\documentclass[twocolumn]{aastex62}
\usepackage{longtable}
\usepackage{comment}

\newcommand{\feh}{\ensuremath{{\rm [Fe/H]}}}
\newcommand{\teff}{\ensuremath{T_{\rm eff}}}
\newcommand{\teq}{\ensuremath{T_{\rm eq}}}
\newcommand{\logg}{\ensuremath{\log{g}}}
\newcommand{\zaspe}{\texttt{ZASPE}}
\newcommand{\ceres}{\texttt{CERES}}
\newcommand{\vsini}{\ensuremath{v \sin{i}}}

\newcommand{\mjup}{\ensuremath{{\rm M_{J}}}}

\newcommand{\mpl}{\ensuremath{{\rm M_P}}}
\newcommand{\rjup}{\ensuremath{{\rm R_J}}}
\newcommand{\rpl}{\ensuremath{{\rm R_P}}}
\newcommand{\rstar}{\ensuremath{{\rm R}_{\star}}}
\newcommand{\mstar}{\ensuremath{{\rm M}_{\star}}}

\newcommand{\rsun}{\ensuremath{{\rm R}_{\odot}}}
\newcommand{\msun}{\ensuremath{{\rm M}_{\odot}}}

\newcommand{\mpkep}{\ensuremath{0.317 \pm 0.026 }}
\newcommand{\rpkep}{\ensuremath{0.833 \pm 0.013 }}
\newcommand{\mskep}{\ensuremath{1.056_{-0.021}^{+0.022} }}
\newcommand{\rskep}{\ensuremath{1.070 \pm 0.010 }}
\newcommand{\per}{\ensuremath{14.893291 \pm 0.000025 }}
\newcommand{\ecc}{\ensuremath{0.476 \pm 0.026 }}

\newcommand{\plname}{K2-287b}
\newcommand{\stname}{K2-287}
\newcommand{\stnameEPIC}{EPIC 249451861}

\graphicspath{{./}{figures/}}

\received{}
\revised{}
\accepted{}
\submitjournal{AAS}

\shorttitle{An Eccentric Warm Saturn}
\shortauthors{Jord\'an et al.}

\begin{document}

\title{\plname: an Eccentric Warm Saturn transiting a G-dwarf}

\correspondingauthor{Andr\'es Jord\'an}
\email{ajordan@astro.puc.cl}

\author[0000-0002-5389-3944]{Andr\'es Jord\'an}
\affiliation{Instituto de Astrof\'isica, Pontificia Universidad Cat\'olica de Chile, Av.\ Vicu\~na Mackenna 4860, Macul, Santiago, Chile}
\affiliation{Millennium Institute for Astrophysics, Chile}

\author[0000-0002-9158-7315]{Rafael Brahm}
\affiliation{Center of Astro-Engineering UC, Pontificia Universidad Cat\'olica de Chile, Av. Vicu\~{n}a Mackenna 4860, 7820436 Macul, Santiago, Chile}
\affiliation{Instituto de Astrof\'isica, Pontificia Universidad Cat\'olica de Chile, Av.\ Vicu\~na Mackenna 4860, Macul, Santiago, Chile}
\affiliation{Millennium Institute for Astrophysics, Chile}

\author[0000-0001-9513-1449]{N\'estor Espinoza}
\altaffiliation{Bernoulli fellow; Gruber fellow}
\affiliation{Max-Planck-Institut f\"ur Astronomie, K\"onigstuhl 17, Heidelberg 69117, Germany }

\author{Cristi\'an Cort\'es}
\affiliation{Departamento de F\'isica, Facultad de Ciencias B\'asicas, Universidad  Metropolitana  de la Educaci\'on, Av. Jos\'e Pedro Alessandri 774, 7760197, Nu\~noa, Santiago, Chile}

\author{Mat\'ias D\'iaz}
\affiliation{Departamento de Astronom\'ia, Universidad de Chile, Camino El Observatorio 1515, Las Condes, Santiago, Chile}

\author{Holger Drass}
\affiliation{Millennium Institute for Astrophysics, Chile}
\affiliation{Center of Astro-Engineering UC, Pontificia Universidad Cat\'olica de Chile, Av. Vicu\~{n}a Mackenna 4860, 7820436 Macul, Santiago, Chile}

\author{Thomas Henning}
\affiliation{Max-Planck-Institut f\"ur Astronomie, K\"onigstuhl 17, Heidelberg 69117, Germany }

\author{James S. Jenkins}
\affiliation{Departamento de Astronom\'ia, Universidad de Chile, Camino El Observatorio 1515, Las Condes, Santiago, Chile}

\author{Mat\'ias I. Jones}
\affiliation{European Southern Observatory, Casilla 19001, Santiago, Chile}

\author[0000-0003-2935-7196]{Markus Rabus}
\affiliation{Center of Astro-Engineering UC, Pontificia Universidad Cat\'olica de Chile, Av. Vicu\~{n}a Mackenna 4860, 7820436 Macul, Santiago, Chile}
\affiliation{Max-Planck-Institut f\"ur Astronomie, K\"onigstuhl 17, Heidelberg 69117, Germany }

\author{Felipe Rojas}
\affiliation{Instituto de Astrof\'isica, Pontificia Universidad Cat\'olica de Chile, Av.\ Vicu\~na Mackenna 4860, Macul, Santiago, Chile}

\author{Paula Sarkis}
\affiliation{Max-Planck-Institut f\"ur Astronomie, K\"onigstuhl 17, Heidelberg 69117, Germany }

\author{Maja Vu\v{c}kovi\'c}
\affiliation{Instituto de F\'isica y Astronom\'ia, Universidad de Vapara\'iso, Casilla 5030, Valpara\'iso, Chile}

\author{Abner Zapata}
\affiliation{Center of Astro-Engineering UC, Pontificia Universidad Cat\'olica de Chile, Av. Vicu\~{n}a Mackenna 4860, 7820436 Macul, Santiago, Chile}

\author{Maritza G.\ Soto}
\affiliation{School of Physics and Astronomy, Queen Mary, University of London, 327 Mile End Road, London, UK}

\author[0000-0001-7204-6727]{G\'asp\'ar \'A.\ Bakos}
\altaffiliation{MTA Distinguished Guest Fellow, Konkoly Observatory, Hungary}
\affiliation{Department of Astrophysical Sciences, Princeton University, NJ 08544, USA}

\author[0000-0001-6023-1335]{Daniel Bayliss}
\affiliation{Department of Physics, University of Warwick, Coventry CV4 7AL, UK}

\author[0000-0002-0628-0088]{Waqas Bhatti}
\affiliation{Department of Astrophysical Sciences, Princeton University, NJ 08544, USA}

\author{Zoltan Csubry}
\affiliation{Department of Astrophysical Sciences, Princeton University, NJ 08544, USA}

\author{Regis Lachaume}
\affiliation{Center of Astro-Engineering UC, Pontificia Universidad Cat\'olica de Chile, Av. Vicu\~{n}a Mackenna 4860, 7820436 Macul, Santiago, Chile}
\affiliation{Max-Planck-Institut f\"ur Astronomie, K\"onigstuhl 17, Heidelberg 69117, Germany }

\author{V\'ictor Moraga}
\affiliation{Instituto de Astrof\'isica, Pontificia Universidad Cat\'olica de Chile, Av.\ Vicu\~na Mackenna 4860, Macul, Santiago, Chile}

\author{Blake Pantoja}
\affiliation{Departamento de Astronom\'ia, Universidad de Chile, Camino El Observatorio 1515, Las Condes, Santiago, Chile}

\author{David Osip}
\affiliation{Las Campanas Observatory, Carnegie Institution of Washington,
Colina el Pino, Casilla 601 La Serena, Chile}

\author{Avi Shporer}
\affiliation{Department of Physics, and Kavli Institute for Astrophysics
and  Space  Research,  Massachusetts  Institute  of  Technology,
Cambridge, MA 02139, USA}

\author[0000-0001-7070-3842]{Vincent Suc}
\affiliation{Instituto de Astrof\'isica, Pontificia Universidad Cat\'olica de Chile, Av.\ Vicu\~na Mackenna 4860, Macul, Santiago, Chile}

\author{Sergio V\'asquez}
\affiliation{Museo Interactivo Mirador, Direcci\'on de Educaci\'on, Av.\ Punta Arenas 6711, La Granja, Santiago, Chile}

\begin{abstract}
We report the discovery of \plname, a Saturn mass planet orbiting a G-dwarf with a period of $P\approx 15$ days. First uncovered as a candidate using K2 campaign 15 data, follow-up photometry and spectroscopy were used to determine a mass of $\mpl=\mpkep\ \mjup$, radius $\rpl=\rpkep\ \rjup$, period $P=\per$ days and eccentricity $e=\ecc$. The host star is a metal-rich $V=11.410 \pm 0.129$ mag G dwarf for which we estimate a mass $\mstar=\mskep$ \msun, radius $\rstar=\rskep$ \rsun, metallicity [Fe/H]=$0.20 \pm 0.05$ and $\teff=5673 \pm 75$ K. This warm eccentric planet with a time-averaged equilibrium temperature of $\teq\approx 800$ K adds to the small sample of giant planets orbiting nearby stars whose structure is not expected to be affected by stellar irradiation. Follow-up studies on the \stname\ system could help in constraining theories of migration of planets in close-in orbits.
\end{abstract}

\keywords{planetary systems -- stars: individual: \stname\ -- planets and satellites: gaseous planets -- planets and satellites: detection}

\section{Introduction}
\label{sec:int}

Giant extrasolar planets that orbit their host stars at distances shorter than $\approx$ 1 AU but farther away than the hot-Jupiter pile-up at $\approx$ 0.1 AU, are termed ``warm" giants. They have been efficiently discovered by radial velocity (RV) surveys \citep[e.g.,][]{hebrard16,jenkins17}, and have a wide distribution for their eccentricities,  with a median of $\approx0.25$. The origin for these eccentricities is a topic of active research because
the migration of planets through interactions with the protoplanetary disc predicts circular orbits \citep{dunhill:2013}, while planet-planet scattering after disc dispersal at typical warm giant orbital distances should generate usually planet collisions rather than high eccentricity excitations \citep{petrovich:2014}.

Transiting giants are key for constraining theories of orbital evolution of exoplanets. Besides providing the true mass of the planet, follow-up observations can be carried out to constrain the sky-projected spin-orbit angle (obliquity) of the system, which is a tracer of the migration history of the planet \citep[e.g.,][]{zhou:2015, esposito:2017, mancini:2018}. While the obliquity for hot giant ($P < 10$ d) systems can be affected by strong tidal interactions \citep{triaud:2013,dawson:2014}, the periastra of warm giants are large enough that significant changes in the spin of the outer layers of the star are avoided, and thus the primordial obliquity produced by the migration mechanism should be conserved.

Unfortunately, the number of known transiting warm giants around nearby stars is still very low. In addition to the scaling of the transit probability as $a^{-1}$, the photometric detection of planets with $P > 10$ days requires
a high duty cycle, which puts strong limitations on the ability of ground-based wide-angle photometric surveys \citep[e.g.,][]{bakos:2004,pollacco:2006,bakos:2013} to discover warm giants. From the total of $\approx 250$ transiting giant planets detected from the ground, only 5 have orbital periods longer than 10 d \citep{kovacs:2010,hatp17,wasp117,brahm:2016:hs17,wasp130}. On the other hand, the \textit{Kepler} and CoRoT space missions found dozens of warm giants \citep[e.g. ][]{corot9,corot10,dawson:2012,borsato:2014}, but orbiting mostly faint stars, for which detailed follow-up observations are very challenging.

Due to their relatively low equilibrium temperatures ($\teq < 1000$ K), transiting warm giants are important objects for characterizing the internal
structure of extrasolar giant planets since their atmospheres are not subject to the yet unknown mechanisms that inflate the radii of typical hot Jupiters \citep[for a review see][]{fortney:2010}.
For warm giants, standard models of planetary structure can be used to infer their internal composition from mass and radii measurements \citep[e.g.,][]{thorngren:2016}. 

In this work we present the discovery of an eccentric warm giant planet orbiting a bright star, having physical parameters similar to those of Saturn. This discovery was made in the context of the K2CL collaboration, which has discovered a number of planetary systems using K2 data \citep{brahm:2016:k2,espinoza:2017:k2,jones:2017,giles:2018,soto:2018,k2-232,k2-261}.

\section{Observations} \label{sec:obs}
\subsection{K2}
\begin{figure*}
\plotone{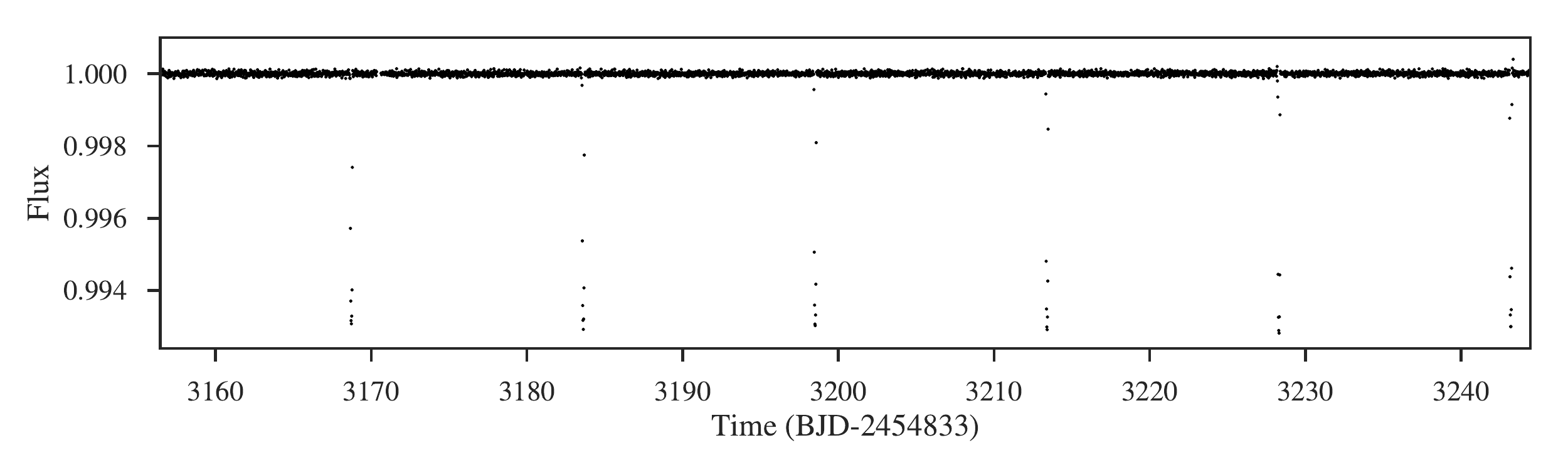}
\caption{De-trended K2 photometry of \stname. Black points are individual 30-min cadence K2 data The transits of \plname\ are clearly seen. \label{fig:lc}}
\end{figure*}

Observations of campaign 15 (field centered at RA=15:34:28 and DEC=-20:04:44) of the K2 mission \citep{howell:2014} took place between August 23 and November 20 of 2017. The data of K2 campaign 15 was released on March 2018.
We followed the steps described in previous K2CL discoveries to process the light curves and identify transiting planet candidates. Briefly, the K2 light curves for 
Campaign 15 were detrended using our implementation of the EVEREST algorithm \citep{luger:2016}, and a Box-Least-Squares \citep[BLS;][]{BLS} algorithm was used to find candidate 
box-shaped signals. The candidates that showed power above the noise 
level were then visually inspected to reject evident eclipsing binary systems 
and/or variable stars. We identified 23 candidates in this field. Among those candidates, \stname\ (\stnameEPIC) stood out as a high priority candidate for follow-up due to its relative long period, deep flat-bottomed transits, and bright host star ($V=11.4$ mag).
The detrended light curves of the six transits observed for \stname\ by K2 are displayed in Figure~ \ref{fig:lc}. 

\subsection{Spectroscopy}

We obtained 52 R=48000 spectra between March and July of 2018 using the FEROS spectrograph \citep{kaufer:99} mounted on the 2.2 MPG telescope in La Silla observatory. Each spectrum achieved a signal-to-noise ratio of $\approx90$ per spectral resolution element. The instrumental drift was determined via comparison with a simultaneous fiber illuminated with a ThAr+Ne lamp. We obtained additionally 25 R=115000 spectra between March and August of 2018 using the HARPS spectrograph \citep{mayor:2003}. Typical signal-to-noise ratio for these spectra ranged  between 30 and 50  per spectral resolution element. 
Both FEROS and HARPS data were processed with the \ceres\ suite of echelle pipelines \citep{brahm:2017:ceres}, which produce radial velocities and bisector spans in addition to reduced spectra.

\begin{figure*}
\plotone{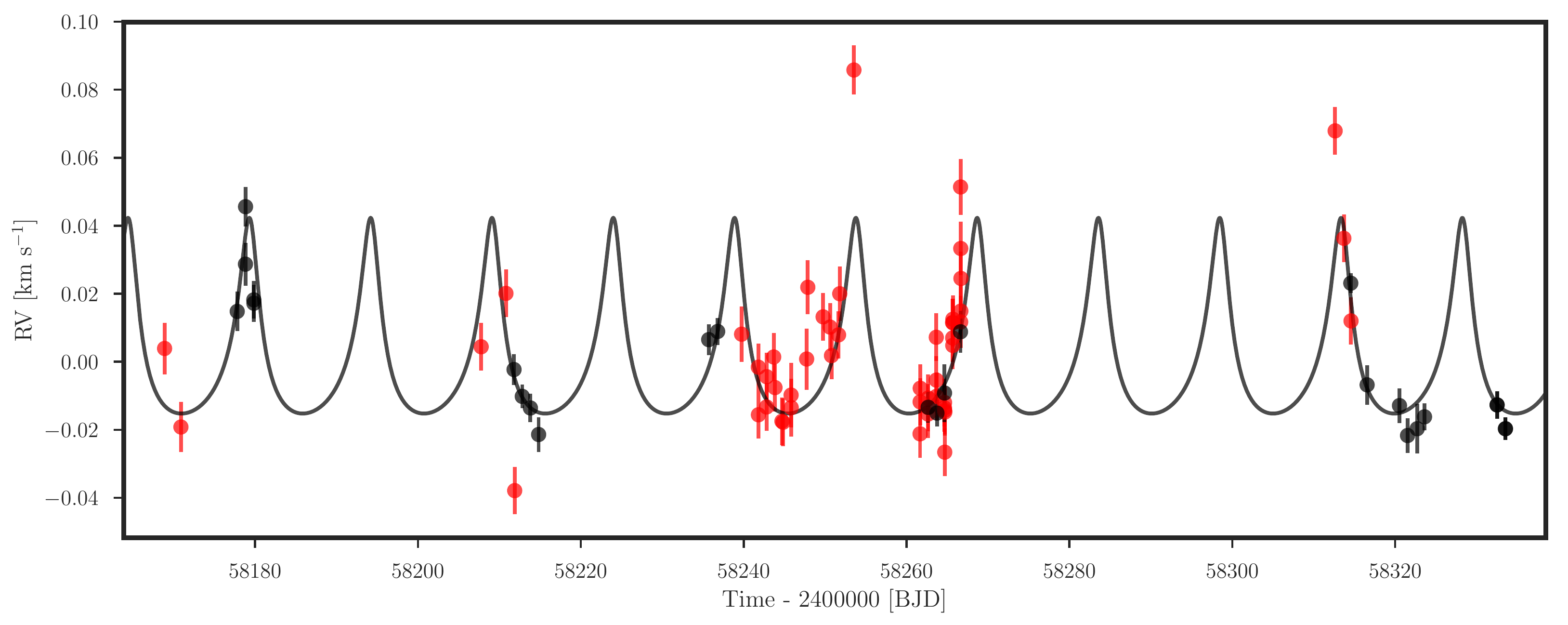}
\caption{Radial velocity (RV) curve for \stname\ obtained with FEROS (red) and HARPS (black). The black line corresponds to the Keplerian model with the posterior parameters found in Section \ref{sec:glob}.\label{fig:rvstime}}
\end{figure*}

\begin{figure}
\plotone{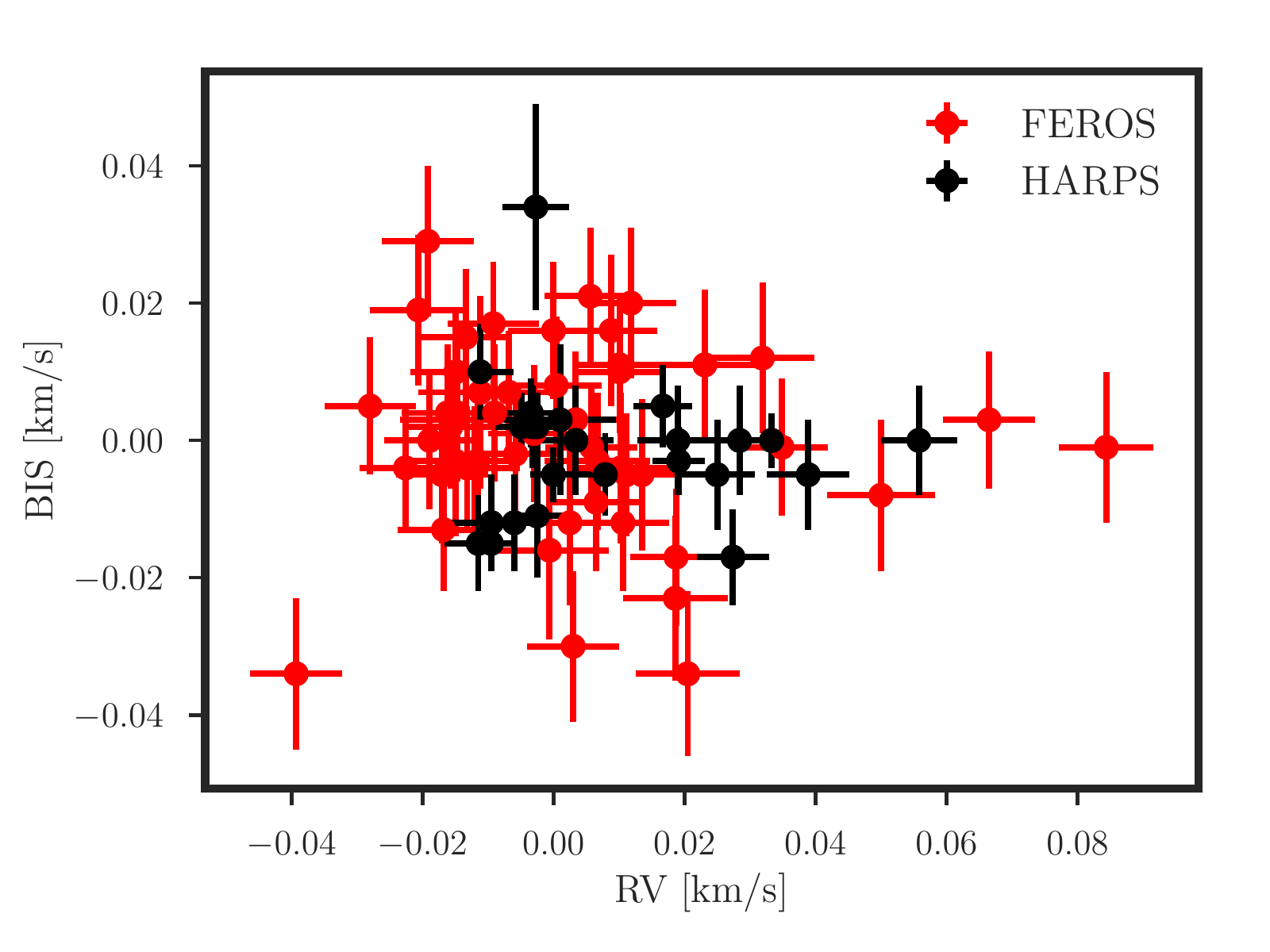}
\caption{Radial velocity (RV) versus bisector span (BIS) scatter plot using data from our spectroscopic observations of \stname. We find that the data is consistent with no correlation. \label{fig:bis}}
\end{figure}

Radial velocities and bisector spans are presented in Table~\ref{tab:rvs} with their corresponding uncertainties, and the radial velocities are displayed as a function
of time in Figure~\ref{fig:rvstime}. No large amplitude variations were identified which could be associated with eclipsing binary scenarios for the \stname\ system and no additional stellar components were evident in the spectra. The radial velocities present a time
correlated variation in phase with the photometric ephemeris, with an amplitude consistent with the one expected to be produced by a giant planet. We find no correlation between the radial velocities and the bisector spans (95\% confidence intervals for the Pearson coefficient are $[-0.19,0.21]$, see Figure~\ref{fig:bis}).

\subsection{Ground-based photometry}
\label{ssec:ground}

On July 14 of 2018 we observed the primary transit of \stname\ with the Chilean-Hungarian Automated Telescope (CHAT), installed at Las Campanas Observatory, Chile. 
CHAT is a newly commissioned 0.7m telescope, built by members of the HATSouth \citep{bakos:2013} team, and dedicated to the follow-up of transiting exoplanets. 
A more detailed account of the CHAT facility will be published at a future date (Jord\'an et al 2018, in prep\footnote{\url{https://www.exoplanetscience2.org/sites/default/files/submission-attachments/poster_aj.pdf}}).
Observations were obtained in the Sloan i' band and the adopted exposure time was of 53 s per image, resulting in a peak pixel
flux for \stname\ of $\approx$ 45000 ADU during the whole sequence. The observations covered
a fraction of the bottom part of the transit and the egress (see Figure~\ref{fig:pht}). The same event was also monitored by one telescope of the Las Cumbres Observatory 1m network \citep{brown:2013:lcogt} at Cerro Tololo Inter-American Observatory, Chile. Observations were obtained with the Sinistro camera with  2mm of defocus in the Sloan i band. The adopted exposure time for the 88 observations taken was 60 s, and reduced images were obtained with the standard Las Cumbres Observatory pipeline (BANZAI pipeline).
The light curves for CHAT and the Las Cumbres 1m telescope were produced from the reduced images using a dedicated pipeline (Espinoza et al 2018, in prep). 

The light curves were detrended by describing the systematic trends as a Gaussian Process with an exponential squared kernel depending on time, airmass and centroid position and whose parameters are estimated simultaneously with those of the transit. A photometric jitter term is also included; this parameter is passed on as a fixed parameter in the final global analysis that determines the planetary parameters (\S~\ref{sec:glob}). In more detail, the magnitude time series is modeled as 

\begin{equation}
m_i = Z + x_1c_{1,i} + x_2 c_{i,2} + \delta_i + \epsilon_i
\end{equation}

\noindent where $Z$ is a zeropoint, $c_1$ and $c_2$ are comparison light curves, $x_1$ and $x_2$ are parameters weighting the light curves, $\delta$ is the transit model and $\epsilon$ is a Gaussian Process to model the noise. The subscript $i$ denotes evaluation at the time $t=t_i$ of the time series. For the Gaussian process, we assume a  kernel given by 

\begin{equation}
k_{ij} = A \exp\left[- \sum_m \alpha_m(x_{m,i} - x_{m,j})^2\right] + \sigma^2\delta_{ij}.
\end{equation}

\begin{figure*}
\plottwo{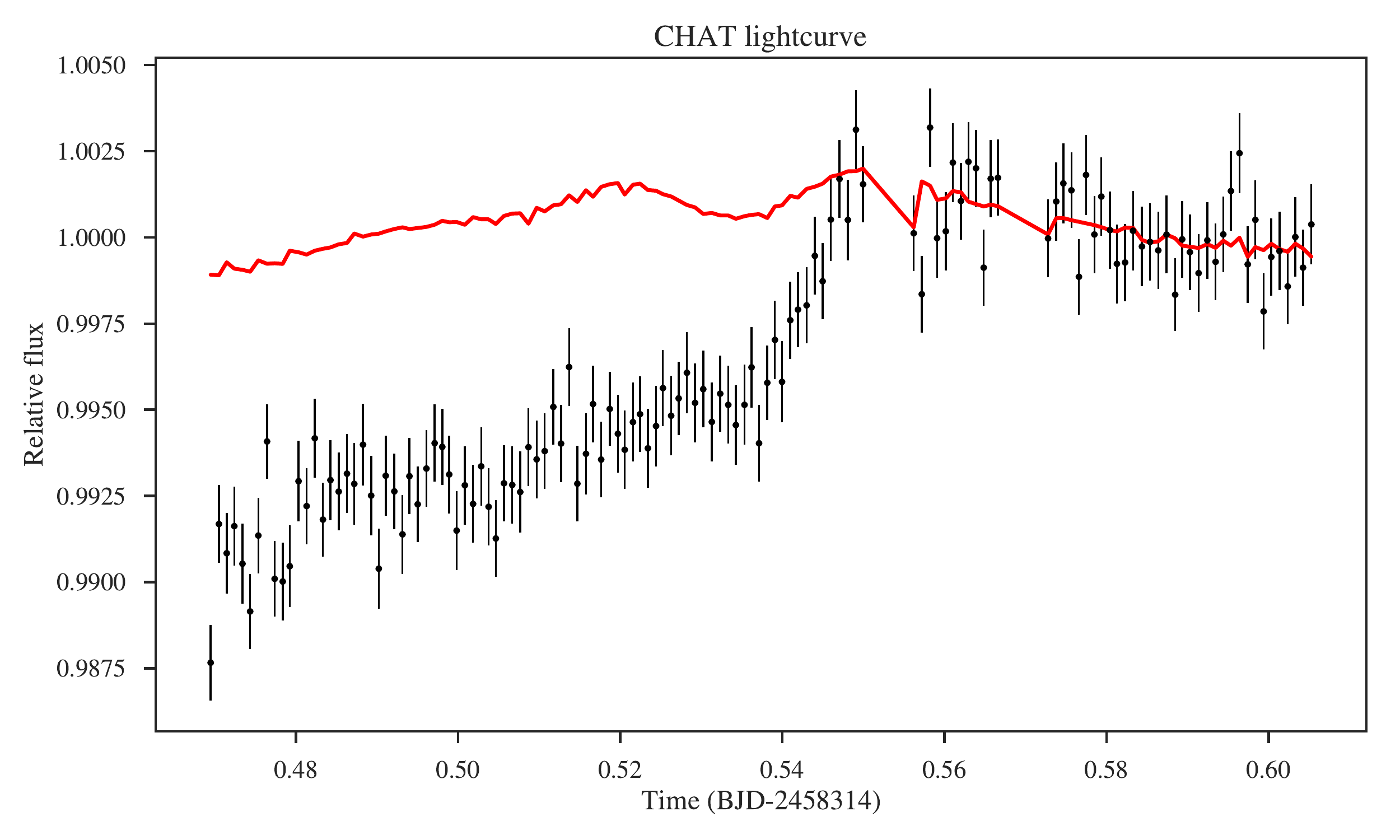}{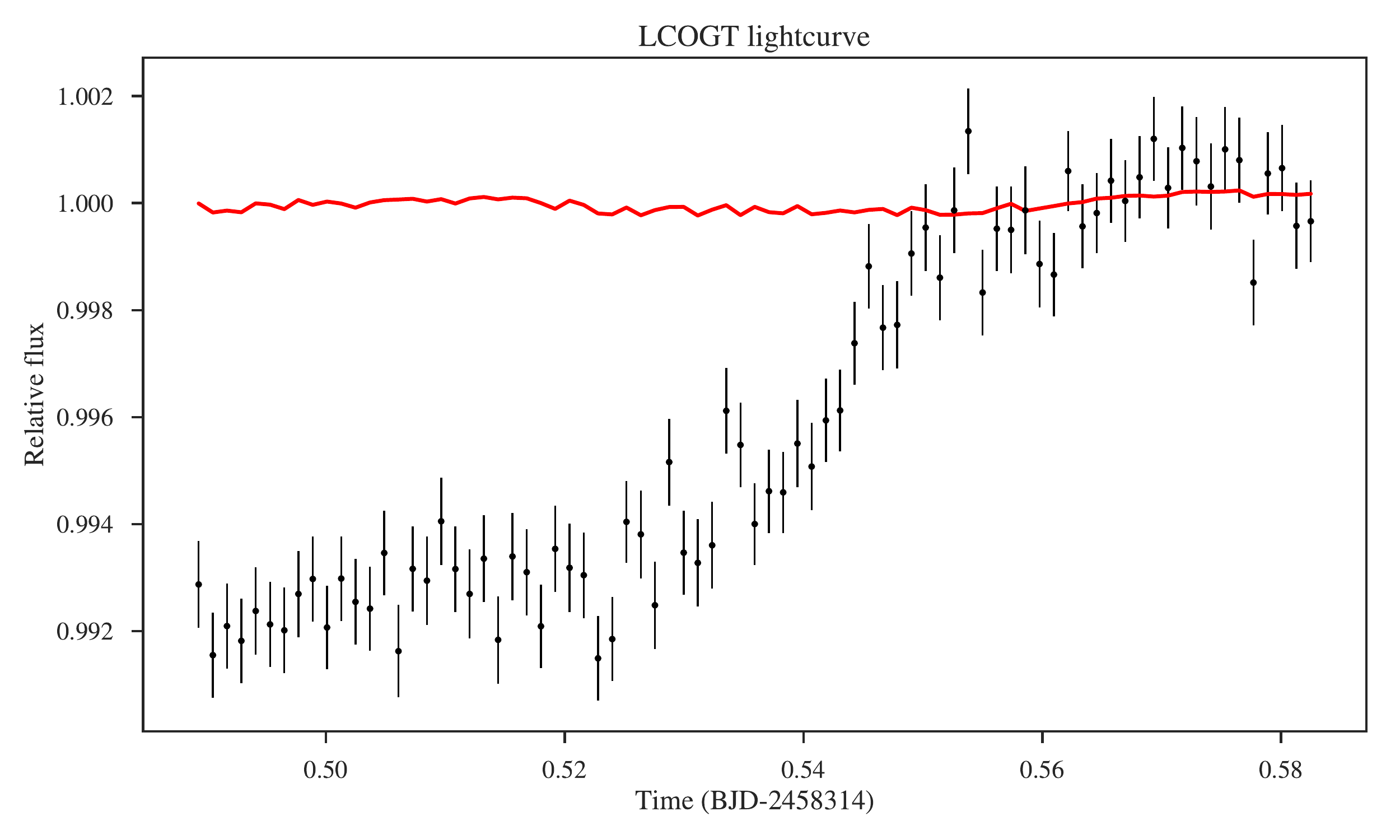}
\caption{Ground-based light curves for the July 14 2018 transit of \plname\ obtained with CHAT (left panel) and a LCOGT 1m telescope at CTIO (right panel). The red lines represent the posterior Gaussian process models for remaining systematics after subtracting the transit and weighted comparison stars and obtained as described in \S\ref{ssec:ground} \label{fig:chat-lcogt}}
\end{figure*}

The variables $x_{m}$ are normalized time ($m=0$), flux centroid in $x$ ($m = 1$) and flux centroid in $y$ ($m=2$); $\delta_{ij}$ is the Kronecker delta. The normalization is carried out by setting the mean to 0 and the variance to 1. The priors on the kernel hyper parameters were taken to be the same as the ones defined in \citet{gibson:2014}, the priors for the photometric jitter term $\sigma$ and $A$ were taken to be uniform in the logarithm between $0.01$ and $100$, with $\sigma$ and $A$ expressed in mmag.
In Figure~\ref{fig:chat-lcogt} we show the CHAT and LCOGT light curves with the weighted comparison stars subtracted along with the Gaussian process posterior model for the systematics.

\subsection{GAIA DR2}

Observations of \stname\ by GAIA were reported in DR2 \citep{gaia, gaia:dr2}. From GAIA DR2, \stname\ has a parallax of $6.29 \pm 0.05$ mas,
an effective temperature of $\teff = 4994 \pm 80$ K and a radius
of $\rstar = 1.18 \pm 0.04 \,\, \rsun$. We used the observed
parallax for \stname\ measured by GAIA for estimating a more precise value of \rstar\ by combining it with the atmospheric parameters obtained from the spectra as described in \S~\ref{sec:ana}. We corrected the GAIA DR2 parallax for the systematic offset of -82 $\mu$as reported in \citet{stassun:2018}.

Two additional sources to \stname\ are identified by GAIA inside the adopted K2 aperture ($\approx 12\arcsec$). However, both stars are too faint ($\Delta G > 7.8$ mag) to produce any significant effect on the planetary and stellar parameters found in \S~\ref{sec:ana}. The radial velocity variations in-phase with the transit signal, which are caused by \stname, confirm that the transit is not caused by a blended stellar eclipsing binary on one of the companions.

\section{Analysis} \label{sec:ana}

\subsection{Stellar parameters}
As in previous K2CL discoveries we estimated the atmospheric parameters of the host star by comparing the co-added high resolution spectrum to a grid of synthetic models through the \zaspe\ code \citep{brahm:2016:zaspe}. In particular, for \stname\ we used the co-added FEROS spectra, because they provide the higher signal-to-noise ratio spectra, and because the synthetic grid of models used by \zaspe\ was empirically calibrated using FEROS spectra of standard stars. Briefly, \zaspe\ performs an iterative search of the optimal model through $\chi^2$ minimization on the spectral zones that are most sensitive to changes in the atmospheric parameters. The models with specific values of atmospheric parameters are generated via tri-linear interpolation of a precomputed grid generated using the ATLAS9 models \citep{atlas9}. The interpolated model is then degraded to match the spectrograph resolution by convolving it with a Gaussian kernel that includes the instrumental resolution of the observed spectrum and an assumed macroturbulence value given by the relation presented in \citet{valenti:2005}. The spectrum is also convolved with a rotational kernel that depends on \vsini, which is considered as a free parameter. The uncertainties in the estimated parameters are obtained from Monte Carlo simulations that consider that the principal source of error comes from the systematic mismatch between the optimal model and the data, which in turn arises from poorly constrained parameters of the atomic transitions and possible deviations from solar abundances.
We obtained the following stellar atmospheric parameters for \stname: \teff=5695 $\pm$ 58 K, \logg=4.4 $\pm$ 0.15 dex, \feh=0.20 $\pm$ 0.04 dex, and \vsini=3.2 $\pm$ 0.2 km s$^{-1}$. The \teff\ value obtained with \zaspe\ is significantly different to that reported by GAIA DR2, but is consistent that of the K2 input catalog \citep{huber:2016}. 

The stellar radius is computed from the GAIA parallax measurement, the available photometry, and the atmospheric parameters. As in \citet{k2-261}, we used a \texttt{BT-Settl-CIFIST} spectral energy distribution model \citep{baraffe:2015} with the atmospheric parameters derived with \zaspe\ to generate a set of synthetic magnitudes at the distance computed from the GAIA parallax. These magnitudes are compared to those presented in table \ref{tab:stprops} for a given value of \rstar. We also consider an extinction coefficient A$_V$ in our modeling which affects the synthetic magnitudes by using the prescription of \citet{cardelli:89}. We explore the parameter space for \rstar\ and A$_V$ using the \texttt{emcee} package \citet{emcee:2013}, using uniform priors in both parameters. We found that \stname\ has a radius of $\rstar=1.07 \pm 0.01$ \rsun\ and has a reddening of A$_V=0.56 \pm 0.03$ mag, which is consistent with what is reported by GAIA DR2.

\begin{figure}
\plotone{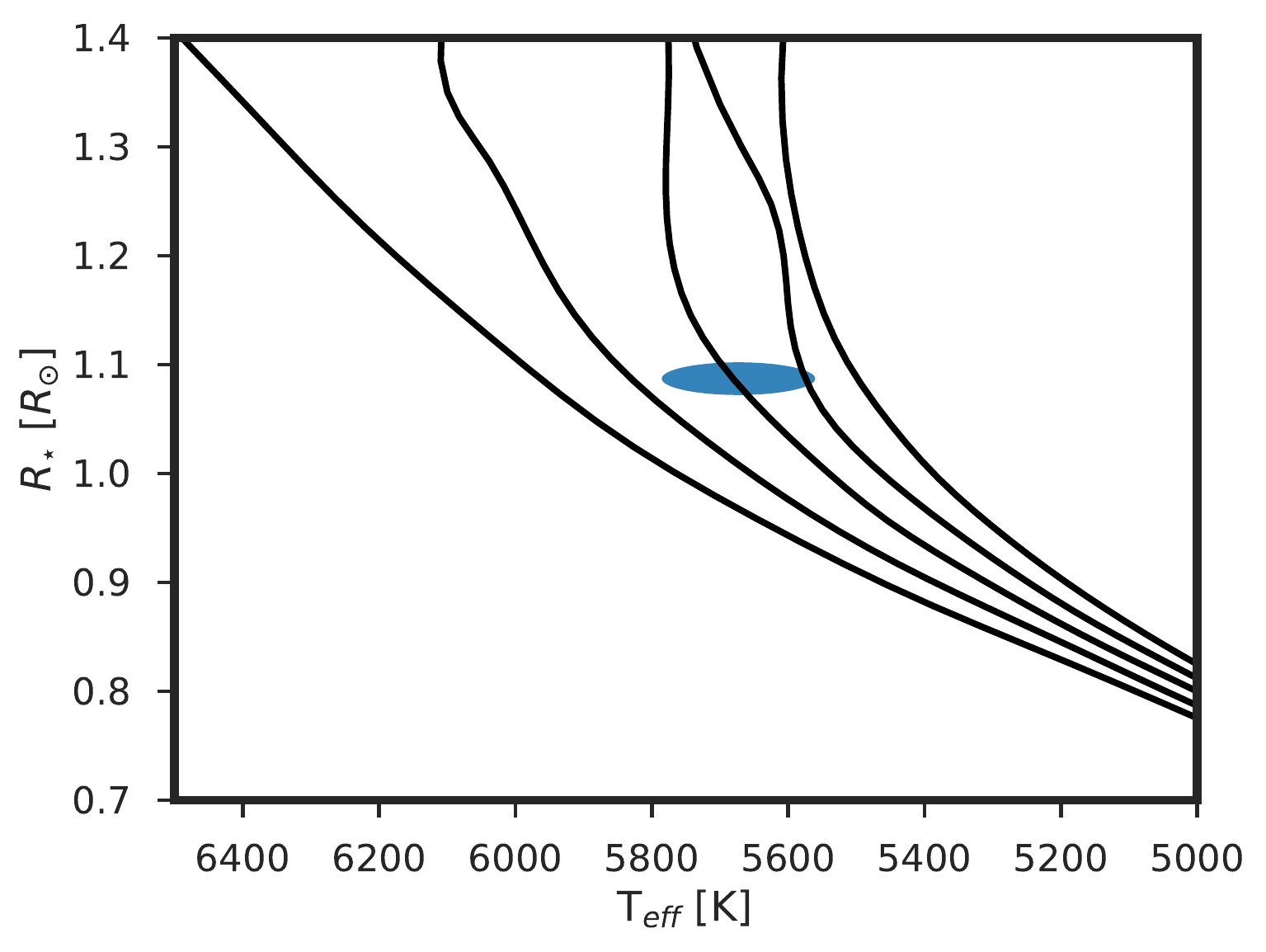}
\caption{Yonsei-Yale isochrones for the metallicity of \stname\ in the \teff--\rstar\ plane. From left to right the isochrones correspond to 1, 3, 5, 7 and 9 Gyr. The position of \stname\  is at the center of the blue shaded region, which marks the 3$\sigma$ confidence region for \teff\ and \rstar.\label{fig:iso}}
\end{figure}

Finally, the stellar mass and evolutionary stage for \stname\ are obtained by comparing the estimation of \rstar\ and the spectroscopic \teff\ with the predictions of the Yonsei-Yale evolutionary models \citep{yi:2001}. We use the interpolator provided with the isochrones to generate a model
with specific values of \mstar, age, and \feh, where \feh\ is fixed to the value found in the spectroscopic analysis. We explore the parameter space for \mstar\ and stellar age using the 
\texttt{emcee} package, using uniform priors in both parameters. We find that the mass and age of \stname\ are $\mstar = 1.036 \pm 0.033$ $\msun$ and 5.6 $\pm$ 1.6 Gyr (see Figure \ref{fig:iso}), similar to those of the Sun. The stellar parameters we adopted for \stname\ are summarized in Table~\ref{tab:stprops}.

\begin{deluxetable*}{lrc}[b!]
\tablecaption{Stellar properties of \stname\  \label{tab:stprops}}
\tablecolumns{3}
\tablewidth{0pt}
\tablehead{
\colhead{Parameter} &
\colhead{Value} &
\colhead{Reference} \\
}
\startdata
Names \dotfill   &    \stname  & EPIC  \\
 & 2MASS J15321784-2221297 & 2MASS  \\
 & TYC 6196-185-1 & TYCHO  \\
 & WISE J153217.84-222129.9 & WISE  \\
RA \dotfill (J2000) &  15h32m17.84s &  EPIC\\
DEC \dotfill (J2000) & -22d21m29.74s &   EPIC\\
pm$^{\rm RA}$ \hfill (mas yr$^{-1}$) & -4.59 $\pm$ 0.11& GAIA\\
pm$^{\rm DEC}$ \dotfill (mas yr$^{-1}$) & -17.899 $\pm$ 0.074 & GAIA\\
$\pi$ \dotfill (mas)& 6.288 $\pm$ 0.051 & GAIA \\ 
\hline
K$_p$ \dotfill (mag) & 11.058 & EPIC\\
B  \dotfill (mag) & 12.009 $\pm$ 0.169 & APASS\\
g'  \dotfill (mag) & 11.727 $\pm$ 0.010 & APASS\\
V  \dotfill (mag) &11.410 $\pm$ 0.129 & APASS\\
r'  \dotfill (mag) & 11.029 $\pm$ 0.010 & APASS\\
i'  \dotfill (mag) & 10.772 $\pm$ 0.020 & APASS\\
J  \dotfill (mag) & 9.677 $\pm$ 0.023 & 2MASS\\
H  \dotfill (mag) & 9.283 $\pm$ 0.025 & 2MASS\\
K$_s$  \dotfill (mag) & 9.188 $\pm$ 0.021 & 2MASS\\
WISE1  \dotfill (mag) & 9.114 $\pm$ 0.022 & WISE\\
WISE2  \dotfill (mag) & 9.148 $\pm$ 0.019 & WISE\\
WISE3  \dotfill (mag) & 9.089 $\pm$ 0.034 & WISE\\
\hline
\teff  \dotfill (K) & 5695 $\pm$ 58& \texttt{zaspe}\\
\logg \dotfill (dex) & 4.398 $\pm$ 0.015 & \texttt{zaspe}\\
\feh \dotfill (dex) & +0.20 $\pm$ 0.04 & \texttt{zaspe}\\
\vsini \dotfill (km s$^{-1}$) & 3.2 $\pm$ 0.2 & \texttt{zaspe}\\
\mstar \dotfill (\msun) & 1.056 $\pm$ 0.022 & YY + GAIA\\
\rstar \dotfill (\rsun) & 1.07 $\pm$ 0.01 & GAIA + this work\\
Age \dotfill (Gyr) & 4.5 $\pm$ 1 & YY + GAIA\\
$\rho_\star$ \dotfill (g cm$^{-3}$) &  1.217 $\pm$ 0.045& YY + GAIA\\
\enddata
\end{deluxetable*}

\subsection{Global modeling}
\label{sec:glob}

\begin{figure*}
\plotone{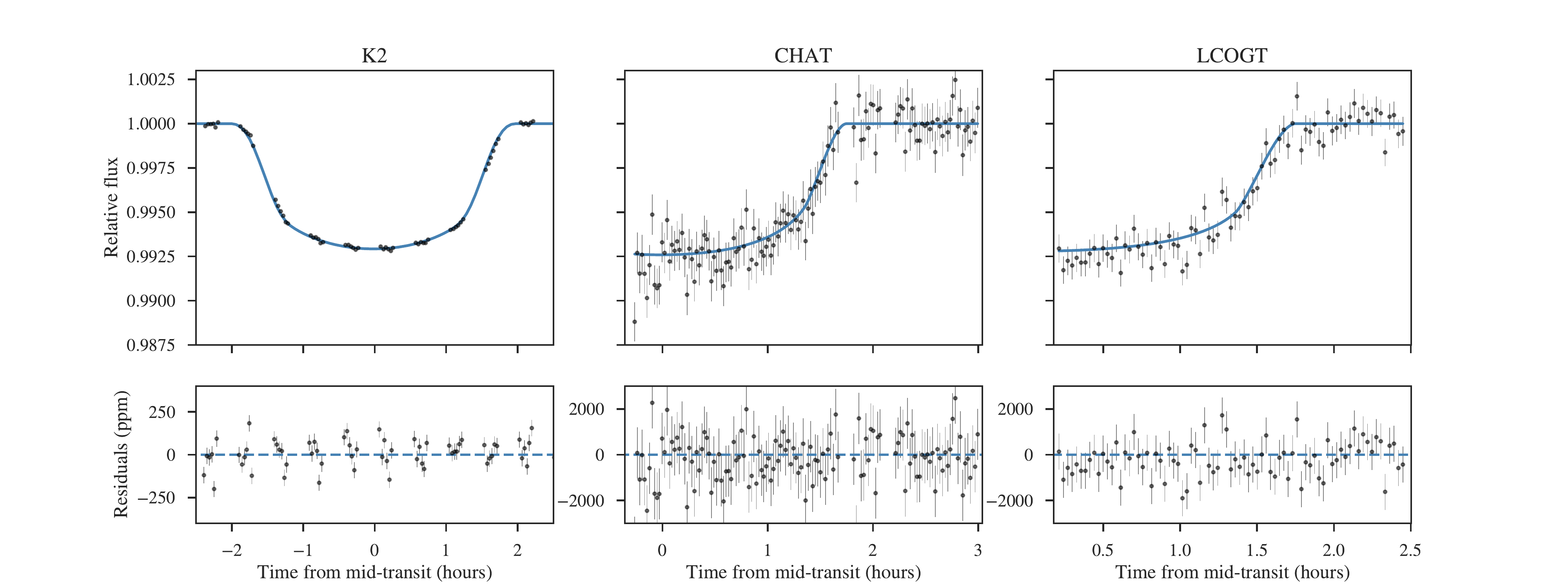}
\caption{The top panels show from left to right: the phase folded Kepler $K2$ photometry ($K_p$ band), the CHAT follow up photometry ($i$ band), and the LCO follow-up photometry ($i$ band) for \stname. For the three cases, the model generated with the derived parameters of \texttt{EXONAILER} is plotted with a blue line. The bottom panels show the corresponding residuals. \label{fig:pht}}
\end{figure*}

\begin{figure*}
\plotone{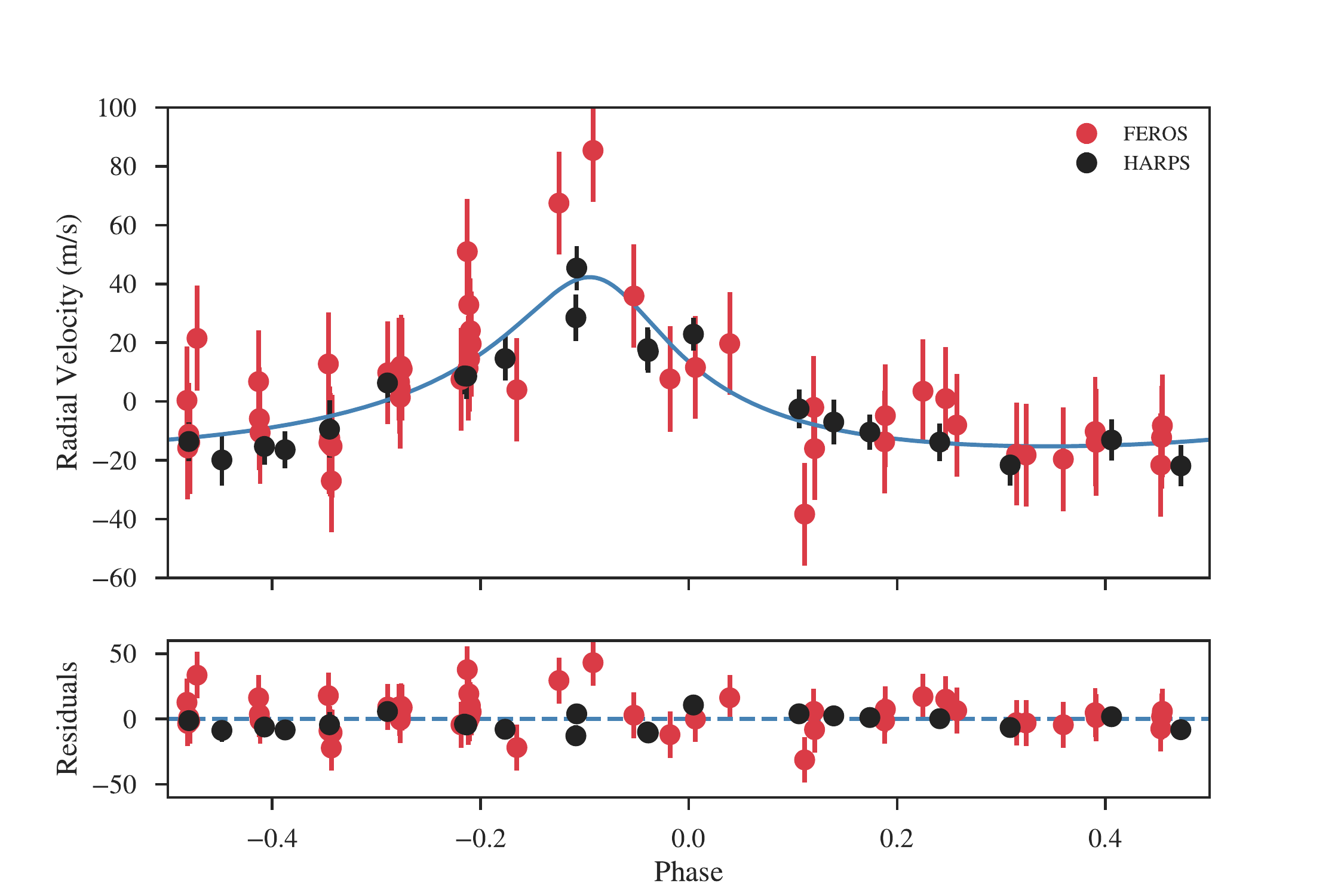}
\caption{The top panel presents the radial velocities for \stname\ (filled circles) obtained with FEROS and HARPS as a function of the orbital phase. The RV model with the derived orbital parameters for \plname\ corresponds to the blue solid line. The bottom panel shows the residuals obtained
 for these radial velocity measurements.
 \label{fig:phr}}
\end{figure*}

In order to determine the orbital and transit parameters of the \plname\ 
system we performed a joint analysis of the detrended K2 photometry, the
 follow-up photometry, and the radial velocities.
As in previous planet discoveries of the K2CL collaboration, we
used the \texttt{exonailer} code which is described in
detail in \citet{espinoza:2016:exo}. Briefly, we model the transit light
curves using the \texttt{batman} package \citep{kreidberg:2015} by taking
into account the effect on the transit shape produced by the long integration time of the 
long-cadence K2 data \citep{ kipping:2010}.
To avoid systematic biases in the determination of the transit parameters
we considered the limb-darkening coefficients as additional free parameters in the transit modeling \citep{EJ:2015}, with the complexity of limb-darkening law chosen following the criteria presented in \citet{espinoza:2016:lds}. In our case, we select the quadratic limb-darkening law, whose coefficients were fit using the 
uninformative sampling technique of \citet{Kipping:LDs}. 
We also include a 
photometric jitter parameter for the K2 data, which allow us to have an estimation of the level of stellar noise in the light curve. The radial velocities are modeled with
the \texttt{radvel} package \citep{fulton:2018}, where we considered systemic velocity and jitter factors for the data of each spectrograph. 
We use the stellar density estimated in our stellar modeling as an extra ``data point" in our global fit as described in \citet{k2-232}. Briefly, there is a term in the likelihood of the form

\begin{eqnarray*}
p(\vec{y}_{\rho_*}|\theta ) = \frac{1}{\sqrt{2\pi \sigma^2_{\rho_*}}} \exp -\frac{(\rho_* - \rho_*^m)^2}
{2\sigma_{\rho_*}^2} ,
\end{eqnarray*}
where 
\begin{eqnarray*}
   \rho^m_* = \frac{3\pi}{GP^2}\left(\frac{a}{R_*}\right)^3
\end{eqnarray*}
by Newton's version of Kepler's law, and 
$\rho_*$ and $\sigma_{\rho_*}$ are the mean stellar density and its standard-deviation,
respectively, derived from our stellar analysis. In essence, because the period $P$ is 
tightly constrained by the observed periodic transits, this extra term puts a strong 
constraint on $a/R_*$, which in turn helps to extract information about the eccentricity $e$ and argument of periastron $\omega$ from the duration of the transit. Resulting planet parameters are set out in Table~\ref{tab:plprops}, the best-fit orbit solution in Figures~\ref{fig:rvstime} and \ref{fig:phr} and the best-fit light curves in Figure~\ref{fig:pht}.

\begin{deluxetable*}{lrc}[b!]
\tablecaption{Planetary properties of the \stname\ system. For the priors, $N(\mu,\sigma)$ stands for a normal distribution with mean $\mu$ and standard deviation $\sigma$, $U(a,b)$ stands for a uniform distribution between $a$ and $b$, and $J(a,b)$ stands for a Jeffrey's prior defined between $a$ and $b$.\label{tab:plprops}}
\tablecolumns{3}
\tablenum{2}
\tablewidth{0pt}
\tablehead{
\colhead{Parameter} &
\colhead{Prior} &
\colhead{Value} \\
}
\startdata
P (days) & $N(14.893,0.01)$  &  14.893291 $\pm$ 0.000025\\
T$_0$ (BJD)&  $N(2458001.722,0.01)$&  2458001.72138 $\pm$ 0.00016 \\
$a$/R$_\star$ & $U(1,300)$ & 23.87$_{-0.31}^{+0.30}$ \\
\rpl/\rstar  & $U(0.001,0.5)$ & 0.08014$_{-0.00098}^{+0.00086}$ \\
$\sigma_w^{\rm K2}$ (ppm) & $J(10,50000)$ & 47.7$^{+0.54}_{0.54}$\\
q$_1^{\rm K2}$ & $U(0,1)$ & 0.32$^{+0.06}_{-0.05}$ \\
q$_2^{\rm K2}$ & $U(0,1)$& 0.57$^{+0.13}_{-0.11}$ \\
q$_1^{\rm CHAT}$ &$U(0,1)$ & 0.83$^{+0.12}_{-0.17}$ \\
q$_2^{\rm CHAT}$ &$U(0,1)$ & 0.15$^{+0.16}_{-0.11}$ \\
q$_1^{\rm LCO}$ & $U(0,1)$& 0.62$^{+0.20}_{-0.19}$ \\
q$_2^{\rm LCO}$ & $U(0,1)$& 0.08$^{+0.11}_{-0.06}$ \\
K (m s$^{-1}$) & $N(0,100)$& 28.8$^{+2.3}_{-2.2}$ \\
$e$ & $U(0,1)$ & 0.478$^{+0.025}_{-0.026}$ \\
$i$ (deg) & $U(0,90)$ & 88.13$^{+0.1}_{-0.08}$\\
$\omega$ (deg) & $U(0,360)$ & 10.1$^{+4.6}_{-4.2}$ \\
$\gamma_{\rm FEROS}$  (m s$^{-1}$)& $N(32963.2,0.1)$& 32930.41$^{+0.10}_{-0.10}$ \\
$\gamma_{\rm HARPS}$ (m s$^{-1}$)& $N(32930.4,0.1)$ & 32963.19$^{+0.10}_{-0.10}$ \\
$\sigma_{\rm FEROS}$ (m s$^{-1}$)& $J(0.1,100)$ & 16.0$^{+2.1}_{-1.8}$ \\
$\sigma_{\rm HARPS}$ (m s$^{-1}$) & $J(0.1,100)$ & 4.8$^{+1.8}_{-1.6}$ \\
\hline
\mpl\ (\mjup)& & 0.315 $\pm$ 0.027\\
\rpl\ (\rjup)& & 0.847 $\pm$ 0.013\\
$a$ (AU)    & & 0.1206$_{-0.0008}^{+0.0008}$\\
\teq\tablenotemark{a}  (K)   & & 804$_{-7}^{+8}$\\ 
\enddata
\tablenotetext{a}{Time-averaged equilibrium temperature computed according to equation~16 of \citet{mendez:2017}}
\end{deluxetable*}

\section{Discussion} \label{sec:dis}

\begin{figure*}
\plotone{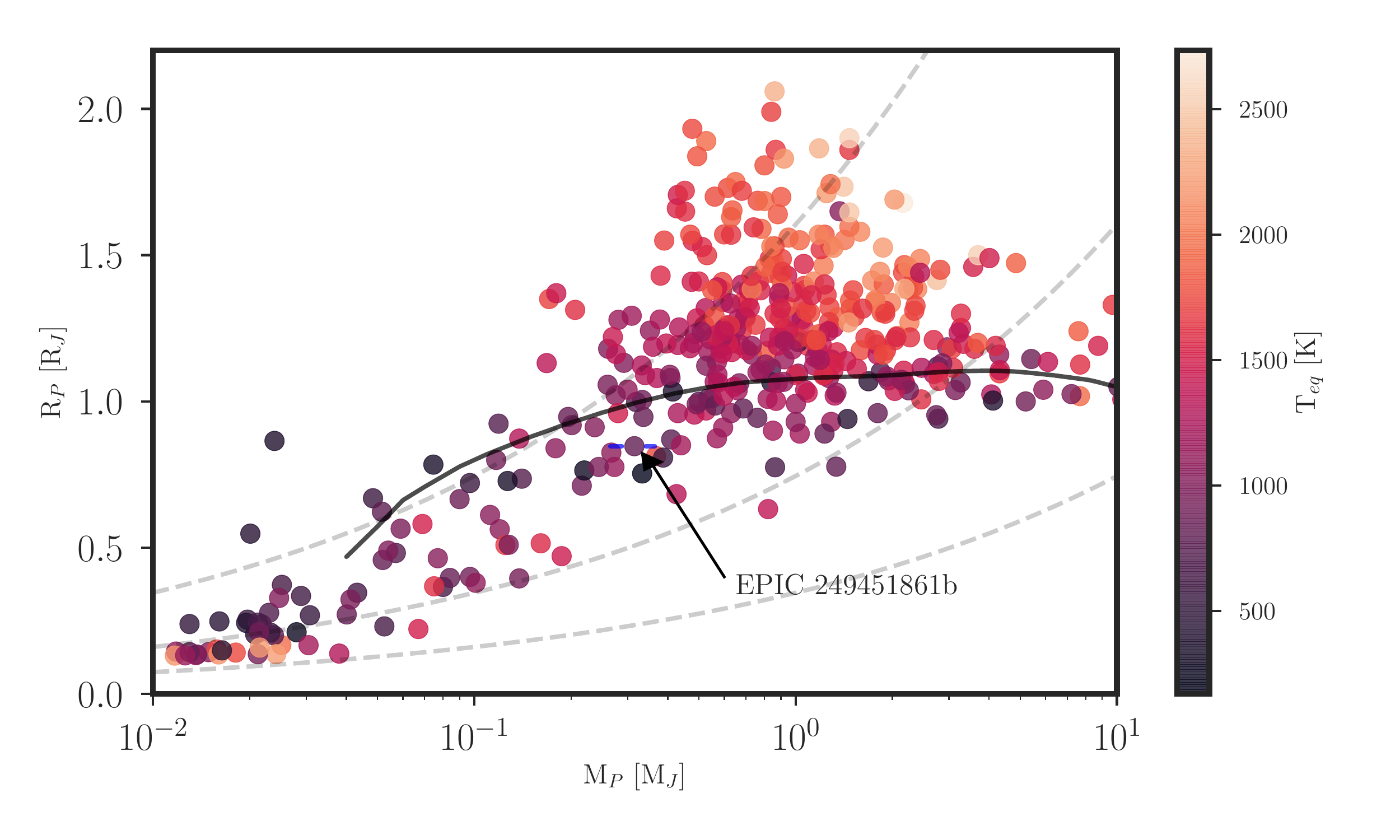}
\caption{Mass-Radius diagram for the full population of transiting planets with both parameters measured to at least 20\% precision. The points are color-coded by equilibrium temperature. \plname\ is the object in the plot that has error bars and is indicated by the arrow. The dashed
gray lines correspond to iso-density curves of 0.3, 3, and 30 g cm$^{-3}$, while the solid
line represents the prediction of the \citet{fortney:2007} structural model with a central core
mass of 10 M$_{\oplus}$.
Due to its relatively low equilibrium temperature, \plname\ lies in a  sparsely populated region of the parameter space of moderately compact giant planets. \label{fig:mr}}
\end{figure*}

By combining data from the Kepler K2 mission and ground based photometry and spectroscopy,
we have confirmed the planetary nature of a $P=14.9$ d candidate around the $V=11.4$ mag G-type
star \stname. We found that the physical parameters of \plname\ (\mpl = \mpkep\ \mjup, \rpl = \rpkep\ \rjup) are consistent to those of Saturn. The non-inflated structure of \plname\ is
expected given its relatively low time-averaged equilibrium temperature of \teq = 808 $\pm$ 8 K.
In Figure \ref{fig:mr} the mass and radius of \plname\ are compared to those for the full population of transiting planets with parameters measured to a precision of 20\% or better. 
Two other transiting planets, orbiting fainter stars, that share similar structural properties to \plname\ are HAT-P-38b \citep{sato:2012} and HATS-20b \citep{bhatti:2016}, which have equilibrium temperatures that are  higher but relatively close to the  $\teq \approx 1000$ K limit below which the inflation mechanism of hot Jupiters does not play a significant role \citep{kovacs:2010,demory:2011}. 
By using the simple planet structural models of \citet{fortney:2007} we find that the observed properties of \plname\ are consistent with having a solid core of $M_c = 31 \pm 4 M_{\oplus}$. However, models that consider the presence of solid material in the envelope of the planet are required to
obtain a more reliable estimate for the heavy element content of \plname\ \citep[e.g.,][]{thorngren:2016}. 

\begin{figure*}
\plotone{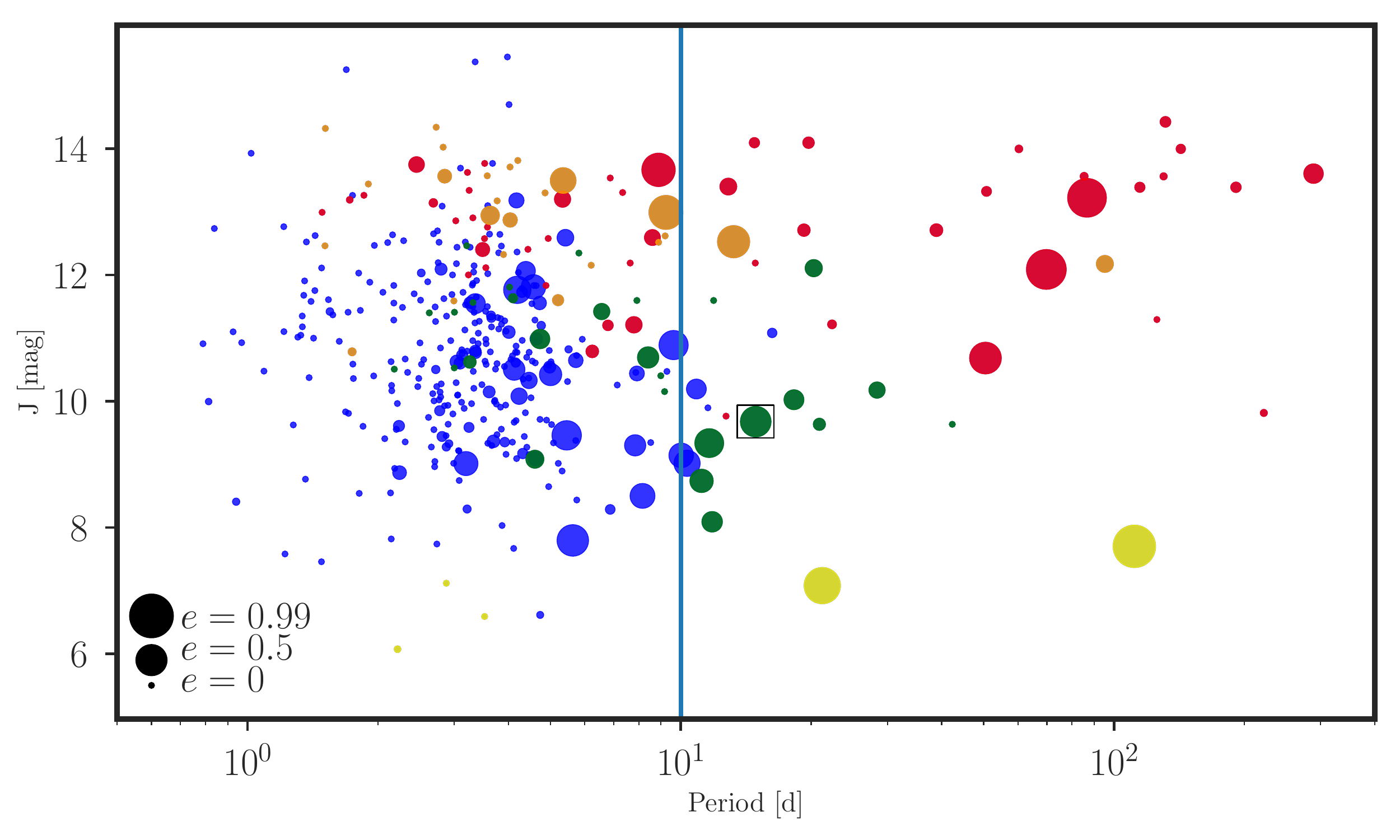}
\caption{Population of well characterized giant planets having \rpl $>$ 0.4 \rjup\ in the orbital period -- J magnitude plane. \plname\ is inside a black square. The size of the points represent the eccentricity of the orbit, while the color indicates the discovery method/mission (blue: ground-based photometry, yellow: RV planets, orange: CoRoT, red: \textit{Kepler}, green: \textit{Kepler} K2). The \textit{Kepler} K2 mission has been the most effective source for discovering transiting bright (J $<$ 11) warm (P $>$ 10 d) giant planets. \label{fig:pj}}
\end{figure*}

The numerous radial velocity measurements obtained for the \stname\ system allow us to constrain the eccentricity of the planet to be $e=0.478 \pm 0.025$. Even though \plname\ is among the most eccentric extrasolar planets to have a period  shorter than 50 days, its periastron distance is not
small enough to cause a significant migration by tidal interactions throughout the main sequence lifetime of the host star. Specifically, by using the equations of \citet{jackson:2009}, we find that in the absence of external sources of gravitational interaction, \plname\ should have possessed an eccentricity of $e\approx0.65$ and a semi-major
axis of $a\approx0.15$ AU when the system was 0.1 Gyr old. Under the same assumptions, we expect that \plname\ would be engulfed by its host star at an age of $\approx$12 Gyr before being able to reach full circularization at a distance of $a\approx0.1$ AU.
These orbital properties for \plname\ and those of the majority of eccentric warm giants are not easy to explain. If \plname\ was formed \textit{in situ} \citep{huang:2016} at 0.15 AU or migrated to this position via interactions with the protoplanetary disc \citep{lin:1997}, its eccentricity could have been excited  by the influence of another massive object in the system after disc dispersal. However, planet-planet scattering \citep{ford:2008} at these close-in orbits  generally produces planet collisions rather than eccentricity excitation \citep{petrovich:2014}.
An alternative proposition for the existence of these eccentric systems is that they
are being subject to secular gravitational interactions produced by another distant
planet or star in the system \citep{rasio:1996}, with the planet experiencing long term
cyclic variations in its eccentricity and spin orbit angle. In this scenario, the planet 
migrates by tidal interactions only during the high eccentricity stages, but it is usually
found with moderate eccentricities. Further observations on the \stname\ system could help
 support this mechanism as the responsible for its relatively high eccentricity, particularly given that \citet{petrovich:2016} concludes that high-eccentricity migration excited by an outer planetary companion can account for most of the warm giants with $e>0.4$. Specifically, long term radial velocity monitoring and the search for transit timing variations could be used to detect the relatively close companions to migrating warm Jupiters proposed by \citet{dong:2014}. Future astrometric searches of companions with GAIA could also be used to find companions and infer the predicted mutual inclination between both orbits, which are predicted to be high \citet{anderson:2017}.

Finally, it is worth noting that an important fraction of the transiting warm giants amenable for detailed characterization ($J<11$ mag) have been discovered in the last couple of years thanks to the  K2 mission (see Figure~\ref{fig:pj}). The combination of relatively long observing campaigns per field, and the increased number of fields monitored, have allowed the discovery and dynamical characterization of several warm giant planets with data from the K2 mission \citep[see Figure~\ref{fig:pj}, ][]{k2-24,k2-99,barragan:2017,shporer:2017,k2-232,k2-234,k2-261,k2-261b}. 
While not particularly designed to discover warm giants, the TESS mission \citep{tess} is expected to discover $\approx$ 120 additional warm giants with $\rpl > 4R_\oplus$ and an incident flux $F < 150 F_\oplus$, where $F_\oplus$ is the incident flux at Earth, around $J\lesssim 11$ mag stars \citep{barclay:2018}. With such population at hand, it will be possible to compare the distributions of eccentricities and obliquities to predictions from different migration mechanisms \citep[e.g. ][]{petrovich:2016} in order to establish a clearer picture about how eccentric warm giant planets originate.

\acknowledgements

A.J.\ acknowledges support from FONDECYT project 1171208, CONICYT project BASAL AFB-170002, and by the Ministry for the Economy, Development, and Tourism's Programa Iniciativa Cient\'{i}fica Milenio through grant IC\,120009, awarded to the Millennium Institute of Astrophysics (MAS). R.B.\ acknowledges support from FONDECYT Post-doctoral Fellowship Project 3180246, and from the Millennium Institute of Astrophysics (MAS).
M.R.D.\ acknowledges support by CONICYT-PFCHA/Doctorado Nacional 21140646, Chile. A.Z.\ acknowledges support by CONICYT-PFCHA/Doctorado
Nacional 21170536, Chile. 
J.S.J.\ acknowledges support by FONDECYT project 1161218 and
partial support by CONICYT project BASAL AFB-170002.
This paper includes data collected by the K2 mission. Funding for the K2 mission is provided by the NASA Science Mission directorate.
This work has made use of data from
the European Space Agency (ESA) mission Gaia (\url{https:
//www.cosmos.esa.int/gaia}), processed by the Gaia Data
Processing and Analysis Consortium (DPAC, \url{ https://www.cosmos.esa.int/web/gaia/dpac/consortium}). Funding for
the DPAC has been provided by national institutions, in particular
the institutions participating in the Gaia Multilateral
Agreement. 
Based on observations collected at the European
Organisation for Astronomical Research in the Southern
Hemisphere under ESO programmes 0101.C-0497, 0101.C-0407, 0101.C-0510.

\vspace{5mm}
\facilities{CHAT~0.7m, LCOGT~1m, MPG~2.2m, ESO~3.6m, \textit{Kepler}, GAIA, APASS, 2MASS, WISE}


\software{EXO-NAILER \citep{espinoza:2016:exo},
          CERES \citep{brahm:2017:ceres,jordan:2014},
          ZASPE \citep{brahm:2016:zaspe,brahm:2015},
          radvel \citep{fulton:2018}
          }


\bibliography{k2clbib}

\appendix
\setcounter{table}{2}
\begin{longtable*}{lrrrrl}
\caption{Relative radial velocities and bisector spans for \stname.\label{tab:rvs}}\\
\hline
\hline
\multicolumn{1}{l}{BJD} & \multicolumn{1}{r}{RV} & \multicolumn{1}{r}{$\sigma_{\rm RV}$} & \multicolumn{1}{r}{BIS} & \multicolumn{1}{r}{$\sigma_{\rm BIS}$} & \multicolumn{1}{l}{Instrument} \\
\multicolumn{1}{l}{\hbox{(2,400,000$+$)}} & \multicolumn{1}{r}{(m s$^{-1}$)} & \multicolumn{1}{r}{(m s$^{-1}$)} & \multicolumn{1}{r}{(m s$^{-1}$)} & \multicolumn{1}{r}{(m s$^{-1}$)} & \multicolumn{1}{l}{} \\
\hline
\endfirsthead
\endhead
\endlastfoot
$ 58168.8957118 $ & $ 32.9339 $ & $ 0.0076 $ & $ -0.033 $ & $ 0.012 $ & FEROS \\
$ 58170.9025854 $ & $ 32.9108 $ & $ 0.0074 $ & $ -0.002 $ & $ 0.011 $ & FEROS \\
$ 58177.8140231 $ & $ 32.9778 $ & $ 0.0058 $ & $ -0.014 $ & $ 0.008 $ & HARPS \\
$ 58178.8260537 $ & $ 32.9917 $ & $ 0.0063 $ & $ -0.014 $ & $ 0.008 $ & HARPS \\
$ 58178.8381972 $ & $ 33.0086 $ & $ 0.0058 $ & $ -0.009 $ & $ 0.008 $ & HARPS \\
$ 58179.8509201 $ & $ 32.9812 $ & $ 0.0055 $ & $ -0.009 $ & $ 0.008 $ & HARPS \\
$ 58179.8616916 $ & $ 32.9802 $ & $ 0.0055 $ & $ -0.026 $ & $ 0.007 $ & HARPS \\
$ 58207.7691953 $ & $ 32.9344 $ & $ 0.0070 $ & $ -0.051 $ & $ 0.011 $ & FEROS \\
$ 58210.8120326 $ & $ 32.9501 $ & $ 0.0070 $ & $ -0.038 $ & $ 0.010 $ & FEROS \\
$ 58211.8033524 $ & $ 32.9607 $ & $ 0.0045 $ & $ -0.014 $ & $ 0.006 $ & HARPS \\
$ 58211.8839384 $ & $ 32.8921 $ & $ 0.0070 $ & $ -0.055 $ & $ 0.011 $ & FEROS \\
$ 58211.8969659 $ & $ 32.8463 $ & $ 0.0083 $ & $ -0.100 $ & $ 0.012 $ & FEROS \\
$ 58212.8162559 $ & $ 32.9528 $ & $ 0.0035 $ & $ -0.014 $ & $ 0.004 $ & HARPS \\
$ 58213.8155843 $ & $ 32.9494 $ & $ 0.0042 $ & $ -0.005 $ & $ 0.005 $ & HARPS \\
$ 58214.8225570 $ & $ 32.9416 $ & $ 0.0051 $ & $  0.001 $ & $ 0.007 $ & HARPS \\
$ 58235.7054437 $ & $ 32.9695 $ & $ 0.0045 $ & $ -0.004 $ & $ 0.006 $ & HARPS \\
$ 58236.8070269 $ & $ 32.9719 $ & $ 0.0040 $ & $ -0.012 $ & $ 0.005 $ & HARPS \\
$ 58239.7443848 $ & $ 32.9381 $ & $ 0.0081 $ & $ -0.024 $ & $ 0.010 $ & FEROS \\
$ 58241.8009423 $ & $ 32.9284 $ & $ 0.0070 $ & $ -0.020 $ & $ 0.010 $ & FEROS \\
$ 58241.8119744 $ & $ 32.9144 $ & $ 0.0070 $ & $ -0.026 $ & $ 0.010 $ & FEROS \\
$ 58242.8136144 $ & $ 32.9167 $ & $ 0.0070 $ & $ -0.017 $ & $ 0.010 $ & FEROS \\
$ 58242.8246191 $ & $ 32.9256 $ & $ 0.0070 $ & $ -0.023 $ & $ 0.010 $ & FEROS \\
$ 58243.6877674 $ & $ 32.9314 $ & $ 0.0070 $ & $ -0.005 $ & $ 0.010 $ & FEROS \\
$ 58243.8443690 $ & $ 32.9224 $ & $ 0.0070 $ & $ -0.017 $ & $ 0.010 $ & FEROS \\
$ 58244.7006355 $ & $ 32.9125 $ & $ 0.0070 $ & $ -0.021 $ & $ 0.010 $ & FEROS \\
$ 58244.8366538 $ & $ 32.9122 $ & $ 0.0070 $ & $  0.008 $ & $ 0.011 $ & FEROS \\
$ 58245.8250104 $ & $ 32.9202 $ & $ 0.0095 $ & $ -0.014 $ & $ 0.014 $ & FEROS \\
$ 58245.8380679 $ & $ 32.9165 $ & $ 0.0085 $ & $ -0.018 $ & $ 0.013 $ & FEROS \\
$ 58247.7318034 $ & $ 32.9308 $ & $ 0.0090 $ & $ -0.037 $ & $ 0.013 $ & FEROS \\
$ 58247.8756418 $ & $ 32.9519 $ & $ 0.0079 $ & $ -0.055 $ & $ 0.012 $ & FEROS \\
$ 58249.7532000 $ & $ 32.9432 $ & $ 0.0070 $ & $ -0.001 $ & $ 0.011 $ & FEROS \\
$ 58250.7827423 $ & $ 32.9318 $ & $ 0.0070 $ & $ -0.013 $ & $ 0.010 $ & FEROS \\
$ 58250.6025575 $ & $ 32.9402 $ & $ 0.0070 $ & $ -0.005 $ & $ 0.011 $ & FEROS \\
$ 58251.6502971 $ & $ 32.9379 $ & $ 0.0070 $ & $ -0.030 $ & $ 0.010 $ & FEROS \\
$ 58251.7959960 $ & $ 32.9500 $ & $ 0.0080 $ & $ -0.044 $ & $ 0.012 $ & FEROS \\
$ 58253.5376199 $ & $ 33.0158 $ & $ 0.0072 $ & $ -0.022 $ & $ 0.011 $ & FEROS \\
$ 58261.6566471 $ & $ 32.9088 $ & $ 0.0070 $ & $ -0.025 $ & $ 0.009 $ & FEROS \\
$ 58261.6676712 $ & $ 32.9182 $ & $ 0.0070 $ & $ -0.025 $ & $ 0.009 $ & FEROS \\
$ 58261.6786827 $ & $ 32.9222 $ & $ 0.0070 $ & $ -0.004 $ & $ 0.009 $ & FEROS \\
$ 58262.6356569 $ & $ 32.9146 $ & $ 0.0070 $ & $ -0.034 $ & $ 0.009 $ & FEROS \\
$ 58262.6501525 $ & $ 32.9193 $ & $ 0.0070 $ & $ -0.025 $ & $ 0.009 $ & FEROS \\
$ 58262.6526217 $ & $ 32.9496 $ & $ 0.0045 $ & $ -0.007 $ & $ 0.006 $ & HARPS \\
$ 58262.6646765 $ & $ 32.9165 $ & $ 0.0070 $ & $ -0.011 $ & $ 0.009 $ & FEROS \\
$ 58263.6490366 $ & $ 32.9372 $ & $ 0.0070 $ & $ -0.022 $ & $ 0.009 $ & FEROS \\
$ 58263.6600382 $ & $ 32.9246 $ & $ 0.0070 $ & $ -0.014 $ & $ 0.009 $ & FEROS \\
$ 58263.6710446 $ & $ 32.9198 $ & $ 0.0070 $ & $ -0.024 $ & $ 0.009 $ & FEROS \\
$ 58263.7327984 $ & $ 32.9479 $ & $ 0.0040 $ & $ -0.007 $ & $ 0.005 $ & HARPS \\
$ 58264.6559473 $ & $ 32.9164 $ & $ 0.0070 $ & $ -0.024 $ & $ 0.011 $ & FEROS \\
$ 58264.6629948 $ & $ 32.9538 $ & $ 0.0085 $ & $ -0.006 $ & $ 0.011 $ & HARPS \\
$ 58264.6669743 $ & $ 32.9180 $ & $ 0.0070 $ & $ -0.006 $ & $ 0.010 $ & FEROS \\
$ 58264.6779962 $ & $ 32.9157 $ & $ 0.0070 $ & $ -0.019 $ & $ 0.009 $ & FEROS \\
$ 58264.6890058 $ & $ 32.9034 $ & $ 0.0070 $ & $ -0.016 $ & $ 0.010 $ & FEROS \\
$ 58264.7000122 $ & $ 32.9152 $ & $ 0.0070 $ & $ -0.017 $ & $ 0.010 $ & FEROS \\
$ 58265.6537546 $ & $ 32.9415 $ & $ 0.0070 $ & $ -0.010 $ & $ 0.010 $ & FEROS \\
$ 58265.6647735 $ & $ 32.9370 $ & $ 0.0070 $ & $  0.000 $ & $ 0.010 $ & FEROS \\
$ 58265.6757786 $ & $ 32.9348 $ & $ 0.0070 $ & $ -0.018 $ & $ 0.010 $ & FEROS \\
$ 58265.6867851 $ & $ 32.9425 $ & $ 0.0070 $ & $ -0.026 $ & $ 0.009 $ & FEROS \\
$ 58265.7022013 $ & $ 32.9415 $ & $ 0.0070 $ & $ -0.011 $ & $ 0.009 $ & FEROS \\
$ 58266.6252665 $ & $ 32.9718 $ & $ 0.0062 $ & $ -0.009 $ & $ 0.008 $ & HARPS \\
$ 58266.6331695 $ & $ 32.9814 $ & $ 0.0082 $ & $ -0.029 $ & $ 0.011 $ & FEROS \\
$ 58266.6441948 $ & $ 32.9417 $ & $ 0.0077 $ & $ -0.025 $ & $ 0.011 $ & FEROS \\
$ 58266.6552239 $ & $ 32.9633 $ & $ 0.0079 $ & $ -0.009 $ & $ 0.011 $ & FEROS \\
$ 58266.6662336 $ & $ 32.9449 $ & $ 0.0078 $ & $ -0.026 $ & $ 0.011 $ & FEROS \\
$ 58266.6772400 $ & $ 32.9545 $ & $ 0.0079 $ & $ -0.010 $ & $ 0.011 $ & FEROS \\
$ 58312.6234698 $ & $ 32.9979 $ & $ 0.0070 $ & $ -0.018 $ & $ 0.010 $ & FEROS \\
$ 58313.6965328 $ & $ 32.9663 $ & $ 0.0070 $ & $ -0.022 $ & $ 0.010 $ & FEROS \\
$ 58314.5467674 $ & $ 32.9861 $ & $ 0.0029 $ & $ -0.009 $ & $ 0.004 $ & HARPS \\
$ 58314.5754726 $ & $ 32.9420 $ & $ 0.0070 $ & $ -0.033 $ & $ 0.010 $ & FEROS \\
$ 58316.5526131 $ & $ 32.9562 $ & $ 0.0058 $ & $ -0.009 $ & $ 0.008 $ & HARPS \\
$ 58320.5251962 $ & $ 32.9501 $ & $ 0.0051 $ & $  0.025 $ & $ 0.015 $ & HARPS \\
$ 58321.5156072 $ & $ 32.9413 $ & $ 0.0051 $ & $ -0.024 $ & $ 0.007 $ & HARPS \\
$ 58322.6976017 $ & $ 32.9433 $ & $ 0.0073 $ & $ -0.021 $ & $ 0.007 $ & HARPS \\
$ 58323.6016488 $ & $ 32.9468 $ & $ 0.0040 $ & $ -0.021 $ & $ 0.007 $ & HARPS \\
$ 58332.5127365 $ & $ 32.9503 $ & $ 0.0040 $ & $ -0.020 $ & $ 0.009 $ & HARPS \\
$ 58333.5353776 $ & $ 32.9433 $ & $ 0.0033 $ & $ -0.021 $ & $ 0.005 $ & HARPS \\
$ 58332.5127365 $ & $ 32.9503 $ & $ 0.0040 $ & $ -0.007 $ & $ 0.005 $ & HARPS \\
$ 58333.5353776 $ & $ 32.9433 $ & $ 0.0033 $ & $ -0.024 $ & $ 0.004 $ & HARPS \\
\hline
\end{longtable*}

\end{document}